# Concurrent Encoding of Frequency and Amplitude Modulation in Human Auditory Cortex: MEG Evidence

Huan Luo[1,2], Yadong Wang[1,2,4], David Poeppel[1,2,4], Jonathan Z. Simon[1,2,3]

[1]Neuroscience and Cognitive Science Program
[2]Department of Biology
[3]Department of Electrical and Computer Engineering
[4]Department of Linguistics
University of Maryland, College Park

Running Head: Simultaneous dynamic envelope and carrier encoding

**Address for Correspondence:**

Huan Luo
Neuroscience and Cognitive Science Program
University of Maryland College Park
1401 Marie Mount Hall
College Park MD 20742
USA

301 405 2587
huanl@wam.umd.edu

**Abstract**

A natural sound can be described by dynamic changes in envelope (amplitude) and carrier (frequency), corresponding to amplitude modulation (AM) and frequency modulation (FM) respectively. Although the neural responses to both AM and FM sounds are extensively studied in both animals and humans, it is uncertain how they are co-represented when changed simultaneously but independently, as is typical for ecologically natural signals. This study elucidates the neural coding of such sounds in human auditory cortex using magnetoencephalography (MEG). Using stimuli with both sinusoidal modulated envelope ($f_{AM}$, 37 Hz) and carrier frequency ($f_{FM}$, 0.3 – 8 Hz), it is demonstrated that AM and FM stimulus dynamics are co-represented in the neural code of human auditory cortex. The stimulus AM dynamics are represented neurally with AM encoding, by the auditory Steady State Response (aSSR) at $f_{AM}$. For sounds with slowly changing carrier frequency ($f_{FM}$ < 5 Hz), it is shown that the stimulus FM dynamics are tracked by the phase of the aSSR, demonstrating neural phase modulation (PM) encoding of the stimulus carrier frequency. For sounds with faster carrier frequency change ($f_{FM} \geq$ 5 Hz), it is shown that modulation encoding of stimulus FM dynamics persists, but the neural encoding is no longer purely PM. This result is consistent with the recruitment of additional neural AM encoding over and above the original neural PM encoding, indicating that both the amplitude and phase of the aSSR at $f_{AM}$ track the stimulus FM dynamics. A neural model is suggested to account for these observations.

**Introduction**

Amplitude modulation (AM) and frequency modulation (FM) are two important physical aspects of communication sounds, corresponding to the independent envelope and carrier dynamics of a sound. They are found in a wide range of species-specific vocalizations for both animals and humans (Doupe & Kuhl, 1999). In speech recognition studies, acoustic envelope (i.e. AM) cues were shown to be crucial to speech intelligibility (Drullman et al., 1994, Shannon et al., 1995). Analogously, Zeng et al (2005) have shown that acoustic carrier (e.g. FM) cues significantly enhance speech recognition performance even under noisy listening conditions, in contrast to AM cues, which enhance recognition only under ideal listening conditions.

Physiological responses to both AM and FM sounds have been widely studied in non-human species (Schreiner & Urbas, 1986, 1988; Eggermont, 1994; Gaese et al., 1995; Heil & Irvine, 1998; Liang et al., 2002), as well as in humans, using electroencephalography (EEG) and magnetoencephalography (MEG) (Rees et al., 1986; Ross et al., 2000; Picton et al., 2003), fMRI (Giraud et al., 2000), and intra-cranial recordings (Liegeois-Chauvel et al., 2004). There is also a rich psychophysical literature of behavioral responses to modulations (Zwicker, 1952; Viemeister, 1979; Moore & Sak, 1996). However, it is still debated whether AM and FM sounds are processed using the same or different mechanisms and pathways (Saberi & Hafter, 1995; Moore & Sek, 1996; Patel & Balaban, 2000, 2004; Dimitrijevic et al., 2001; Liang et al., 2002). Animal studies show that cortical neurons can fire phase-locked to amplitude modulated sounds up to tens of Hertz (Schreiner & Urbas, 1986, 1988; Eggermont, 1994; Gaese et al., 1995). However, rate coding instead of temporal coding has been observed for higher rates (Lu et al., 2001). In addition, there is a high degree of similarity between cortical responses to AM and FM stimuli (Liang et al., 2002), suggesting at least some shared representation of temporal modulations by cortical neurons (Wang et al., 2003). Correspondingly, in EEG and MEG studies with human subjects, auditory steady-state responses (aSSR) at the modulation frequency were found for both AM (Ross et al., 2000; Rees et al., 1986) and FM sounds (Picton et al., 2003), consistent with the stimulus-synchronized discharge (or the temporal coding) observed in animal studies. In one MEG experiment, Ahissar et al. (2001), using speech stimuli with very complex envelopes, showed that the first principle component of the recorded signal was correlated with the speech stimulus envelopes (AM). Cumulatively, these results reveal that cortex apparently encodes incoming auditory signals by decomposing them into envelope and carrier (Smith et al., 2002).

Natural sounds, however, contain simultaneously modulated envelope and carrier frequencies (both AM and FM). Therefore, instead of manipulating the envelope or carrier dynamics separately, the auditory cortex may be probed using stimuli with both dynamic envelope and carrier. Elhilali et al. (2004) have shown that single units from primary auditory cortex (AI) in ferrets lock to both slow AM and FM modulations and to the fast fine structure of the carrier (up to carrier frequencies of a few hundred Hz). In humans, Dimitrijevic et al. (2001) employed independent amplitude and frequency modulation (IAFC) stimuli with relatively higher modulation frequencies (above 80 Hz) and found independent aSSR responses for both AM and FM using EEG. Patel & Balaban (2000, 2004), using MEG, investigated the processing of sinusoidally amplitude modulated tone *sequences* (co-modulation of both envelope and carrier where the slow frequency modulation is periodic but not sinusoidal), and showed that the phase of the aSSR at the envelope modulation frequency tracks the tone sequences, i.e. the carrier changes. This indicates a relation between the representation of dynamic changes in envelope and carrier in human auditory cortex. For complex stimuli, such as these, it is not clear whether the envelope and carrier dynamics are generally represented independently, or are co-represented, at least at some stage of auditory cortical processing.

How might auditory cortex co-represent envelope and carrier dynamics simultaneously? Modulation encoding is one important possibility. Modulation is a way to describe stimulus dynamics, such as the AM and FM signals, it is also a very important method to embed a general information-bearing signal into a second signal, or to co-represent two signals. AM, FM, and related modulation schemes are widely used encoding techniques in both nature and electrical engineering. One class of modulation encoding is AM, in which the modulation signal is used to modulate the amplitude of another signal, called the carrier. Another important class is phase modulation (PM), in which the signal needing to be transmitted modulates the phase of the carrier signal. FM is a generalized PM, in which the signal needing to be transmitted modulates the time derivative of the carrier phase (which is also equal to the carrier's instantaneous frequency). These encoding schemes can be used to transmit signals even in the presence of noise, whether electromagnetically in the radio band, or neurally in the auditory system (Oppenheim & Willsky, 1997). Figure 1a illustrates these basic concepts from the engineering encoding point of view. Figure 1b shows the hypothesized spiking activity corresponding to neural modulation encoding (third row: PM encoding; fourth row: AM encoding) of the considered stimulus with sinusoidally modulated carrier frequency (first row, FM) and amplitude (second row, AM). An ensemble of PM encoding neurons (third row) will produce an evoked neural PM signal similar to that shown in the middle of the lower panel of Figure 1a (obtained mathematically by low-pass filtering the spike train). Similarly, an ensemble of AM encoding neurons (fourth row) will produce an evoked neural AM signal similar to that shown in the middle of the upper panel of Figure 1a. This neural modulation encoding model will be addressed in more detail in the Discussion.

Figure 1 about here

In the Fourier domain, modulated signals have distinctive signatures, which may be easier to detect and decode than their time-domain versions. A narrowband carrier appears as a single peak in the spectrum at $f_{carrier}$, the carrier frequency. The modulations due to either pure AM or pure PM appear as sideband frequency patterns in the spectrum. Specifically, the spectrum will have an upper sideband at $f_{carrier} + f_{modulation}$ and a lower sideband at $f_{carrier} - f_{modulation}$ (often accompanied by additional, lower power, sidebands at more distant frequencies). At least one example of modulation encoding is seen in human auditory cortex: at extremely slow frequency modulations (~0.1 Hz), the phase of the envelope modulation frequency aSSR tracks the carrier change, i.e. a form of PM encoding (Patel & Balaban, 2000, 2004). Whether other methods are used, and what method is used at higher frequencies, is largely unknown.

The ability of auditory cortex to track stimulus dynamics via the aSSR is limited. The aSSR to AM sounds can be recorded with MEG from humans at stimulus rates up to ~100 Hz, with a large peak around 40 Hz (Ross et al., 2000); EEG responses follow to higher rates (see, e.g. Picton et al., 2003) but responses at those higher rates are not generated by auditory cortex. The aSSR at the modulation frequency, however, is generated only by neural temporal coding, whereas many neurons employ rate coding for rapidly modulated stimuli (Lu et al., 2001). Therefore, it is still not fully understood how - and how fast - auditory cortex can track a stimulus, particularly for stimuli modulated in both envelope and carrier, as is typical of most ecologically relevant signals.

 The present study was designed to address three questions: First, how does human auditory cortex represent or co-represent simultaneous AM and FM. Second, how fast can human auditory cortex track the carrier dynamics (FM). Third, is there any coding transition as the rate of carrier dynamics increases? To address these issues, we take advantage of the high temporal resolution of MEG, which has shown to be a method with outstanding sensitivity to record from human auditory cortex.

**Materials and Methods**

*Subjects*

12 subjects (8 males) with normal hearing and no neurological disorders provided informed consent before participating in this experiment. The subjects' mean age was 25 and all were right handed. A digitized head shape was obtained for each subject for use in equivalent-current dipole source estimation.

*Stimuli*

Nine stimuli were created, using custom-written MATLAB programs (The MathWorks, Natick, MA), with a sampling frequency of 44.1 kHz. The stimuli were sinusoidally frequency modulated tones with modulation frequencies ($f_{FM}$) of 0.3, 0.5, 0.8, 1.0, 1.7, 2.1, 3.0, 5.0, 8.0 Hz and frequency deviation between 220 Hz to 880 Hz. In addition, the entire stimulus amplitude was modulated sinusoidally at a fixed rate of 37 Hz ($f_{AM}$) with modulation depth of 0.8. All stimuli were 10 s in duration and shaped by rising and falling 100 ms cosine squared ramps. Each stimulus was presented 10 times. Figure 2 shows the spectrogram (higher panel), the spectrum (middle panel) and the temporal waveform (lower panel) of example stimuli, confirming that the stimulus sounds contain both sinusoidally modulated temporal envelope at $f_{AM}$ (37 Hz) and sinusoidally modulated carrier frequency at $f_{FM}$ (0.8 Hz and 2.1 Hz as examples drawn here). Because the frequency range of the carrier ranges from 220 Hz to 880 Hz, the stimuli have the broadband spectra shown in middle panel.

Figure 2 about here

To ensure that subjects attend to the long stimulus sequences, 36 distracter stimuli were created and inserted into the experiment for subjects to detect. Those distracters were the same as the normal stimuli except single short-duration FM sweeps were inserted at random time in the stimulus. Subjects were instructed to press a button when detecting the distracter stimuli. Normal stimuli (90 = 9 × 10) and distracter stimuli (36) were mixed and played in a pseudo-random order at a comfortable loudness level to subjects. Subjects performed the required task fairly well (average miss rate: ~ 3/36; average false alarm rate: ~ 1/36). The entire experiment was divided into 4 blocks with breaks between them. Only the data for normal stimuli were further analyzed.

*MEG recordings*

Neuromagnetic signals were recorded continuously with a 157 channel whole-head MEG system (5 cm baseline axial gradiometer SQUID-based sensors, KIT, Kanazawa, Japan) in a magnetically shielded room, using a sampling rate of 1000 Hz and an online 100 Hz analog low-pass filter, with no high-pass filtering. Each subject's head position was determined via five coils attached to anatomical landmarks (nasion, left and right pre-auricular points, two forehead points) at the beginning and the end of recording to ensure that head movement was minimal. Head shape was digitized using a three-dimensional digitizer (Polhemus, Inc.).

*Data analysis*

Data from 10 trials for a given condition (same $f_{FM}$) were concatenated (total of 100 s per condition) and were discrete Fourier transformed (DFT) using 100000 points. DFT was performed on data of all 157 MEG channels, and for all 9 stimulus conditions.

*Phasor representation and channel selection*

For each channel, the steady state response (aSSR) at 37 Hz ($f_{AM}$) is parameterized by the DFT component's magnitude and phase at 37 Hz ($f_{AM}$). The result is a map of complex aSSR, i.e. a map of complex magnetic field values. An example of such a map can be seen in Figure 3b, where the complex magnetic field at each channel is represented by a phasor, i.e. an arrow with length proportional to the complex field magnitude and with direction given by the complex field phase (Simon and Wang, 2005). The 10 channels per subject with the largest magnitudes across all the channels in both hemispheres at the 37 Hz ($f_{AM}$) modulation frequency were regarded as channels representative of auditory cortical activity and selected for further analysis, motivated by the positive relationship between tracking performance and response strength at $f_{AM}$ found in an MEG experiment exploring representation of tone sequence in human auditory cortex (Patel and Balaban, 2004).

*aSSR and M100 Equivalent-Current Dipole localization*

To localize the neural source of the aSSR, the complex aSSRs corresponding to $f_{FM} = 0.3$ Hz were analyzed to determine the best (least mean square) fit for a pair of equivalent-current dipoles (Simon and Wang, 2005). The resulting complex dipoles' positions, one in each hemisphere, are the estimates of the source locations. These aSSR source locations are compared to the M100 source locations, estimated by the purely real version of the same algorithm. The M100 was measured in a pretest experiment, in which subjects were instructed to count the number of 1 kHz pure tones they heard. The M100 component is believed to originate in the superior temporal cortex on the upper bank of the superior temporal gyrus slightly posterior to Heschl's gyrus on the planum temporale (Lutkenhoner and Steinstrater, 1998). This direct comparison permits an analysis of the aSSR location without requiring magnetic resonance image (MRI).

*Sideband confusion matrix*

To test for the presence of general modulation encoding, including the possibility of AM and PM encoding, we examined the spectra of the MEG responses to co-modulated stimuli for a two-sideband pattern: with strong spectral peaks at $f_{AM} \pm f_{FM}$, a distinctive signature of modulation encoding.

Target sideband frequencies were defined for different $f_{FM}$ as upper sideband ($f_{AM} + f_{FM}$) and lower sideband ($f_{AM} - f_{FM}$), leading to 18 (9 × 2) frequencies (upper: 37.3, 37.5, 37.8, 38, 38.7, 39.1, 40, 42, 45 Hz; lower: 36.7, 36.5, 36.2, 36, 35.3, 34.9, 34, 32, 29 Hz). The DFT amplitude and phase at every target sideband frequency were extracted for 10 channels (selected specifically per subject), for every stimulus condition, giving an 18 × 9 × 10 × 12 data set (frequency × stimulus_condition × channel × subject).

Confusion matrix analysis was used to assess statistical significance. In this methodology, any one particular sideband frequency is examined for *all* stimulus conditions (even those whose responses should not elicit the sideband). Ideally, the response to the one stimulus whose frequency modulation is at the corresponding frequency examined in the response should elicit higher magnitude at that frequency than those of all other stimuli. Then, even under noisy conditions, at a particular target sideband frequency, more channels should elicit the highest magnitude for the stimulus condition with the appropriate $f_{FM}$ than any other stimulus condition. For example, for the target sideband frequency of 38 Hz (37+1, the upper sideband for stimulus with $f_{FM}$ = 1 Hz), the stimulus with $f_{FM}$ = 1 Hz should elicit a larger number of channels with maximum strength at 38 Hz than any of the other stimuli (other $f_{FM}$). If that is true, we claim that modulation encoding is employed to co-represent the envelope and carrier dynamics characterized by $f_{AM}$ = 37 Hz and $f_{FM}$ = 1 Hz.

For each target sideband frequency, the magnitudes at this frequency for all 9 stimulus conditions were compared and the stimulus which elicited maximum magnitude at this sideband frequency was stored, indexed by its stimulus condition (out of 9). This calculation was performed for all target sideband frequencies (9 upper sidebands and 9 lower sidebands), for all the 10 selected channels, giving an 18 × 10 × 12 (frequency × channel × subject) analysis set. Each cell represents the index number of the stimulus condition inducing maximum response at this frequency, for each channel and each subject. Because it is possible that only one of the two sideband frequencies was detectable in the MEG signal due to different signal to noise ratios, upper and lower sideband frequencies were explored separately. For each subject two separate 9 × 9 confusion matrixes were constructed to represent the upper and lower sideband performance. For example, in upper sideband confusion matrix (Figure 5a), columns represent stimuli conditions ($f_{FM}$ of 0.3 Hz to 8 Hz) and rows represent different target upper sideband frequencies (37 + 0.3 to 37 + 8 Hz). Each element in the matrix represents the number of channels that were the largest magnitude elicited at this frequency (corresponding row) by this stimulus (corresponding column). The sum each one row is equal to 10 (channels) and thus each row actually reflects the histogram of stimulus condition that drove the specific sideband frequency most across 10 channels. Ideally, if every sideband frequency is maximally elicited by the corresponding stimulus condition, the confusion matrix will be purely diagonal.

We construct upper and lower sideband confusion matrices for each subject, and also, to represent the total sideband performances, the sum of the two confusion matrixes (upper panels of each of the subfigures in Figure 5 show the sum of the confusion matrix across 12 subjects). To further determine whether sidebands were significantly elicited (i.e. whether the diagonal is significantly peaked in the whole confusion matrix) and to explore differences for different stimulus conditions, the results from the diagonal axis were extracted from both upper and lower sideband confusion matrix for each subject. These 9-element arrays, were normalized (to range from 0 to 1), and correspond to the proportion of channels which showed the correct maximum sideband for each subject (see the lower panels of each of the subfigures in Figure 5 for the grand averages across 12 subjects). For example, a value of 1 means that, for that target sideband frequency, the stimulus with the corresponding $f_{FM}$ elicited maximum magnitude for all channels; whereas a value of 0.2 reflects that only 20% of all the selected channels showed maximum magnitude at this target sideband frequency when the corresponding stimulus occurred. The same procedure was also applied in the total sideband confusion matrix where the data range was normalized to range from 0 to 2 so it roughly shows whether two or one sidebands were elicited. A Monte Carlo simulation was used to calculate the 95% significance threshold for the proportion value for both the one-sideband confusion matrix and the total sidebands confusion matrix (dotted-starred line in the lower panels of each of the subfigures in Figure 5 and Figure 6). The same confusion matrix procedure was used to investigate direct aSSR at $f_{FM}$ frequencies (0.3 – 8 Hz) and shown in Figure 5d.

*Simulations of confusion matrix performance*

The nine-element diagonals of the three confusion matrices in the lower panels of Figures 5a, 5b, and 5c are measures of sideband performance. They are used to determine the statistical significance of modulation encoding for different stimulus dynamics, specifically, the different FM ($f_{FM}$, 0.3 – 8 Hz). A simulation was performed to compare the confusion matrix performance and sideband performance for pure PM encoding with the empirical results. Only the simulation of pure PM encoding is shown, but pure AM encoding would provide similar results. Confusion matrix performance by itself cannot distinguish between the types of modulation encoding, and it is only one way to check the possibility of modulation encoding. But, by comparing the simulation results with real MEG results, we are informed as to whether modulation encoding is employed at all.

A simulation of neural responses with pure PM encoding was created with neural carrier frequency 37 Hz ($f_{AM}$), neural modulation frequencies ($f_{FM}$) of 0.3 – 8 Hz, random starting phases ($\phi_1, \phi_2$), and neural modulation depth of 0.6. The simulation signals were created by adding Gaussian white noise (*GWN*), the level of which was adjusted to match the real neural sideband performance:

$$S_{PM} = \cos(2\pi f_{AM} t + \phi_1 + 0.6\cos(2\pi f_{FM} t + \phi_2)) + GWN$$

The simulated signal results in a confusion matrix, just as for the empirical data. Block simulations represented 12 subjects, composed of the 9 different $f_{FM}$ conditions), each of which was simulated 10 times (representing 10 channels). Then the confusion matrix for the higher sideband, the lower sideband, and the total sideband performance (the same frequencies as in empirical data) was calculated using the same procedures described above. These are shown in the upper panels of each of the subfigures of Figure 6. The sideband performance for each of the 3 confusion matrixes was extracted from the diagonal of corresponding simulated confusion matrix. These are shown in the lower panels of each of the subfigures of Figure 6.

*Encoding-type parameter calculation*

Sidebands naturally occur for all types of modulation coding (including AM and PM). To help determine which modulation coding created the sidebands, an encoding-type parameter ($\alpha$, defined below, ranging between 0 and $2\pi$) was calculated to distinguish AM encoding from PM encoding. Both encoding mechanisms (see Figure 1b) elicit two sidebands, but with different phase relationships across the sidebands and carrier.

As will be seen below, AM encoding produces $\alpha$ near 0 (or $2\pi$); PM encoding produces $\alpha$ near $\pi$ (for reasonably moderate phase modulation index values). The encoding-type parameter $\alpha$ is defined as $(\theta_{upper} - \theta_{AM}) - (\theta_{AM} - \theta_{lower})$, using, respectively, the phase at the sidebands $f_{upper} = f_{AM} + f_{FM}$, $f_{lower} = f_{AM} - f_{FM}$, and carrier $f_{AM}$.

The mathematical derivation follows. For neural response carrier frequency $f_c$ (identified with $f_{AM}$), neural response modulation frequency $f_m$ (identified with $f_{FM}$), and modulation index $m$, this is shown for the neural response case of AM:

$$\begin{aligned}
S_{AM}(t) &= (1 + m\cos(2\pi f_m t + \phi_1))\cos(2\pi f_c t + \phi_2) \\
&= \cos(2\pi f_c t + \phi_2) + \frac{m}{2}\cos(2\pi(f_c + f_m)t + \phi_1 + \phi_2) + \frac{m}{2}\cos(2\pi(f_c - f_m)t + \phi_2 - \phi_1) \\
&= \cos(2\pi f_c t + \phi_2) + \frac{m}{2}\cos(2\pi f_{upper} t + \theta_{upper}) + \frac{m}{2}\cos(2\pi f_{lower} t + \theta_{lower})
\end{aligned}$$

Where we have set $\theta_{upper} = \phi_1 + \phi_2$ and $\theta_{lower} = \phi_2 - \phi_1$. Thus,

$\alpha_{AM} := (\phi_{upper} - \phi_2) - (\phi_2 - \phi_{lower}) = ((\phi_1 + \phi_2) - \phi_2) - (\phi_2 - (\phi_2 - \phi_1)) = 0$, which is also equivalent to $\alpha_{AM} = 2\pi$.

Correspondingly in the neural PM case,

$$\begin{aligned}
S_{PM}(t) &= \cos(2\pi f_c t + \phi_3 + m\cos(2\pi f_m t + \phi_4)) \\
&= \cos(2\pi f_c t + \phi_3)\cos(m\cos(2\pi f_m t + \phi_4)) - \sin(2\pi f_c t + \phi_3)\sin(m\cos(2\pi f_m t + \phi_4)) \\
&\approx \cos(2\pi f_c t + \phi_3) - m\sin(2\pi f_c t + \phi_3)\cos(2\pi f_m t + \phi_4) \\
&\approx \cos(2\pi f_c t + \phi_3) + \frac{m}{2}\cos(2\pi f_{upper} t + \phi_3 + \phi_4 + \frac{\pi}{2}) + \frac{m}{2}\cos(2\pi f_{lower} t + \phi_3 - \phi_4 + \frac{\pi}{2}) \\
&\approx \cos(2\pi f_c t + \phi_3) + \frac{m}{2}\cos(2\pi f_{upper} t + \theta_{upper}) + \frac{m}{2}\cos(2\pi f_{lower} t + \theta_{lower})
\end{aligned}$$

Where we have set $\theta_{upper} = \phi_3 + \phi_4 + \frac{\pi}{2}$ and $\theta_{lower} = \phi_3 - \phi_4 + \frac{\pi}{2}$ giving,

$$\alpha_{PM} = (\theta_{upper} - \phi_3) - (\phi_3 - \theta_{lower}) = ((\phi_3 + \phi_4 + \frac{\pi}{2}) - \phi_3) - (\phi_3 - (\phi_3 - \phi_4 + \frac{\pi}{2})) = \pi,$$

concluding the mathematical derivation.

Experimentally, the encoding-type parameter α may take either of these values or any value between, and so a distribution of measured values is expected. α was calculated for all 9 different $f_{FM}$ stimuli conditions, all 10 selected channels and all 12 subjects. A histogram of α distribution across channels and subjects was drawn for each $f_{FM}$ stimulus condition.

It should be noted that the calculation presented for $\alpha_{PM}$ is only valid for small modulation index $m$ (found to be smaller than π/4 by Patel and Balaban, 2004), but it can be shown numerically that the result is robust even for moderately large values of $m$ (up to ~3π).

Encoding-type parameter statistics

Circular statistics were used to estimate the (circular) mean and (circular) standard error of α. To calculate the circular mean value $\bar{\alpha}$, for each $f_{FM}$, all the α were first converted into complex vectors ($e^{i\alpha}$) and the mean of those complex vectors was determined. The circular mean $\bar{\alpha}$ is the four-quadrant inverse tangent of this complex vector mean. The circular standard error of α ($SE_\alpha$) was calculated using bootstrap (balanced, 1000 instances) across the α of all the selected channels and all the 12 subjects. (Efron & Tibshirani, 1994; Fisher, 1996)

*Simulations of mixed neural PM encoding and AM encoding*

A simulation was performed to see how different neural encoding schemes using mixed AM encoding and PM encoding affect the resulted α parameter distribution. The simulation results are compared with the empirical α distribution data and provide suggestions for possible mechanisms for sidebands appearance in real MEG data (e.g., pure AM encoding, pure PM encoding, or mixture of AM encoding and PM encoding).

Simulated pure neural AM encoding signals and PM encoding signals with carrier frequency of 37 Hz ($f_{AM}$) and modulation frequency of 2 Hz (one example of $f_{FM}$) were created with random starting phase ($\phi_i$) and the simulation mixture signals were created by combining them using different weights $\tau$.

$$S_{AM}(t) = (1 + 0.4\cos(2\pi f_{FM} t + \phi_1))\cos(2\pi f_{AM} t + \phi_2)$$

$$S_{PM}(t) = \cos(2\pi f_{AM} t + \phi_3 + \cos(2\pi f_{FM} t + \phi_4))$$

$$S(t) = \tau S_{AM}(t) + (1-\tau) S_{PM}(t).$$

The encoding-type parameter α for this simulated signal was then calculated as above. We performed 1000 simulations for each weight parameter $\tau$ that ranged from 0.1 to 0.9 in steps of 0.1 and calculated the α distribution histogram for different values of $\tau$.

**Results**

*Auditory steady-state response at $f_{AM}$*

Figure 3a shows the discrete Fourier transform of one channel of a representative subject, including the aSSR at $f_{AM}$ (37 Hz). The spectrum shows a clear peak at 37 Hz, the AM frequency $f_{AM}$. Because of the limited signal-to-noise ratio in the MEG signal, other peaks (external narrowband noise) are also observable (and known to be not due to movement or related artifacts, or from bad sensors). The relevance of using sidebands to detect neural modulation coding is that the vast majority of the noise peaks cannot interfere with the sidebands. Figure 3b shows the corresponding phasor representations for aSSR at 37 Hz for all channels (Simon and Wang, 2005). There is a clear bilateral auditory cortical origin for aSSR at 37 Hz. Figure 3c shows the grand average results for both the aSSR equivalent-current dipole (red) and the M100 (green). The dipole locations of aSSR and of M100 activity were compared across all subjects, and it was found that they have displacement not significantly different from 0 (for right hemisphere: $\Delta x$ = -1.1±5.3 mm, $\Delta y$ = 4.6±7.6 mm, $\Delta z$ = -2.4±5.8 mm; for left hemisphere: $\Delta x$ = -0.0±3.2 mm, $\Delta y$ = 4.4±8.2 mm, $\Delta z$ = -4.1±5.4 mm). This result supports the idea that the source of aSSR is in superior temporal cortex since the M100 component is believed to originate there (Lutkenhoner and Steinstrater, 1998). This result is consistent with the aSSR localization results of Ross et al. (2000) given the resolution limitations of this data set.

Figure 3 about here

*Auditory steady-state response at sidebands*

Figure 4 shows the aSSR at upper sidebands for the same channel in the same subject at different stimulus conditions. First, the aSSR at 37 Hz ($f_{AM}$) can be seen for all 9 different stimulus conditions (black arrow); Secondly, stimuli with specific $f_{FM}$ elicited corresponding sidebands (here, only upper sidebands are shown, grey arrows; the lower sidebands, not shown, do not necessarily follow the same pattern). For example, for stimulus $f_{FM} = 0.5$ Hz, the response at 37.5 Hz (= 37 + 0.5) is elicited, and when stimulus $f_{FM} = 1$ Hz, the response at 38 Hz (= 37 + 1) is elicited. For this one channel, the upper sideband for $f_{FM}$ of 5 Hz is not visible. Note that narrowband noise coexists with the sidebands we want to detect.

Figure 4 about here

*Sideband performance*

Figure 5 shows the sum of confusion matrixes across all subjects. We can see that for both upper and lower sideband confusion matrices (Figure 5a,b), most rows peak on the diagonal, reflecting that the stimulus did strongly elicit responses at the upper and lower sideband frequencies. Figure 5c is the sum of upper and lower confusion matrix across all subjects and also clearly shows the peaks along the diagonal. The curve below each confusion matrix is the corresponding diagonal value vector, and the starred line is the 95% threshold. The total sideband performance (Figure 5c) is well above the threshold for all the stimuli we tested here. There is some difference between upper and lower sideband performance (Figure 5a,b). Specifically, the poor performance in the upper sideband for the two lowest values of $f_{FM}$ is artifactual, due to strong narrowband system noise at the corresponding upper sideband frequencies (37.3, 37.5 Hz), but present for almost all channels and all subjects. The narrowband noise at those two frequencies can be seen for all 9 conditions in Figure 4, and clearly masks any elicited sidebands at those frequencies.

Figure 5 about here

Figure 6 about here

*Simulation of confusion matrix and sideband performance*

The simulation results (Figure 6abc) can be compared with the experimental results (Figure 5abc), to demonstrate to what extent that modulation encoding is be employed. As can be seen in Figures 5a and 6a, the upper sideband performance of real MEG data matches well with the simulation (except for $f_{FM}$ of 0.3 Hz and 0.5 Hz, which was discussed above, can arise as an artifact due to narrowband noise at 37.3 Hz and 37.5 Hz). The empirical lower sideband performance matches well with the simulated lower sideband performance (Figures 5b, 6b) for $f_{FM}$ below 5 Hz. Considering upper and lower sideband performances together, as reflected in empirical total sideband performance (Figure 5c, 6c), we can confirm that modulation encoding is used for the entire $f_{FM}$ range tested here (0.3 – 8 Hz). The deteriorated performance for lower sideband performance for $f_{FM}$ above 5 Hz may be due to some kind of encoding transition, but because the performance for the upper sideband is still above threshold during that range (Figure 5a), this demonstrates that *some* form of modulation encoding is present (even if not pure PM or pure AM encoding).

*Auditory steady-state response at $f_{FM}$*

The significance of the responses at the $f_{FM}$ (i.e. not at the corresponding sidebands of $f_{AM}$) was explored using the same confusion matrix procedure. Figure 5d shows the confusion matrix for the actual $f_{FM}$ frequencies (not sidebands elicited around $f_{AM}$). and we can see that most of the stimuli, especially the stimuli with higher $f_{FM}$ (> 0.5 Hz) showed aSSR at corresponding $f_{FM}$ frequency.

*Encoding-type parameter $\alpha$*

Figure 7a shows the $\alpha$ histograms for different $f_{FM}$. For lower $f_{FM}$ (< 5 Hz), the $\alpha$ distribution is peaked and centered near or at $\pi$ (the PM encoding region), except at 0.3 Hz. For the highest $f_{FM}$ (5 Hz, 8 Hz), $\alpha$ shows a more uniform-like distribution between 0 and $2\pi$. In addition, using circular statistics, the mean and standard error of the encoding-type parameter $\alpha$ are shown in Figure 7b for different $f_{FM}$. The gray bars define the PM encoding region, $\pi \pm \pi/4$, and AM encoding region, within $\pi/4$ of 0 or $2\pi$ (the range is arbitrary and for illustrative purposes only). For the lower $f_{FM}$ range ($f_{FM}$ < 5 Hz, except at 0.3 Hz), the encoding-type parameter $\alpha$ is near $\pi$ and within the PM encoding region. As $f_{FM}$ increases, $\alpha$ begins to leave the PM encoding region, but at the same time becomes more uniformly distributed and the bootstrap derived circular error of the mean becomes larger. The uniform-like distribution for $f_{FM}$ of 0.3 Hz is also explained by the narrowband noise at the upper sideband frequency (37.3 Hz), which in turn leads to a noisier encoding parameter distribution.

Figure 7 about here

*Simulation of α-dependence on neural AM and PM encoding mixtures*

As stated in the introduction, the spectral sideband can arise from a variety of modulation encodings, including AM encoding: the amplitude of aSSR at $f_{AM}$ (37Hz) tracking the carrier frequency change. A simulation demonstrates whether additional involvement of AM encoding can account for the observed α distribution for higher $f_{FM}$ ( > 5 Hz). Figure 8 shows the α distribution for different mixtures. As can be seen, when the AM encoding contribution is very small (e.g., $\tau = 0.1$), so that the coding is dominated by PM encoding, α is narrowly distributed around π. When the AM encoding contribution α is increased and thus the signal is a more balanced mixture of AM encoding and PM encoding, α approaches a more uniform distribution ($\tau = 0.5, 0.6$). When the AM encoding contribution $\tau$ is large (e.g., $\tau = 0.9$), the signal is dominated by AM encoding, and α peaks around 0 (or 2π).

Comparing the simulation results with our experimental results, we see that the α distribution for lower $f_{FM}$ (< 5 Hz) is similar to the simulation results with small AM encoding weight $\tau$ (0.1–0.3), although the simulation has a narrower distribution. This supports a model of PM encoding dominance at lower $f_{FM}$ rates. Interestingly, in our results for higher $f_{FM}$ (5 Hz, 8 Hz), the α distribution is more uniform, which looks like the simulations with AM encoding and PM encoding mixed in similar proportions. This suggests that the experimental results for higher $f_{FM}$ may be due to involvement of additional AM encoding.

Figure 8 about here

**Discussion**

Human auditory cortex encodes a sound's envelope dynamics (AM), as well as its carrier frequency dynamics (FM). To investigate the way auditory cortex represents different carrier dynamics, we used a specifically designed acoustic stimulus, a sinusoidally co-modulated stimulus with fixed envelope dynamics ($f_{AM}$ =37 Hz) and varying the carrier dynamics. We explored the possibility that auditory cortex co-represents the envelope and carrier dynamics simultaneously using modulation encoding by determining whether a spectral sideband pattern is elicited. In addition, by changing the carrier dynamics from slow to fast (0.3 Hz to 8 Hz), we investigated the possibility of a coding transition (PM encoding vs. AM encoding).

*Relationship to previous aSSR findings*

Consistent with previous research (Ross et al., 2000), we find a robust aSSR at $f_{AM}$ (37 Hz here), which means auditory cortex demodulates the incoming sound and extracts the envelope. The aSSR at $f_{FM}$ is consistent with EEG studies using pure frequency modulated stimuli (Picton et al., 1987), which is one way auditory cortex represents pure carrier dynamics, although they tested much higher modulation frequencies (>80 Hz) than those used here. Dimitrijevic et al. (2001), used independent amplitude and frequency modulation (IAFM) stimuli with also higher modulation frequencies and found separate AM and FM aSSR responses that are relatively independent of each other, suggesting separate and independent encoding of envelope and carrier. We also found the aSSR at $f_{FM}$, but since our AM frequency was fixed, we cannot estimate whether the aSSR at $f_{AM}$ and $f_{FM}$ were independent of each other. When the source of the aSSR was localized using equivalent-current dipoles, no significant difference was found between the location of these dipoles and those of the (well-studied) M100.

*Sidebands and modulation encoding*

Spectral sideband patterns were found throughout our results, either in the upper sideband or lower sideband confusion matrix, indicating that auditory cortex does use modulation encoding to co-represent envelope and carrier dynamics simultaneously. The detection of the spectral sideband pattern alone, however, does not determine the particular type of modulation encoding (e.g. PM vs. AM). Note that the stimuli employed here to probe the cortical response differ only in FM rates, from slow to moderately fast, sharing all other properties: common spectral widths, envelope dynamics (37Hz), and temporal structure (simultaneous AM and FM), as shown in Figure 2. Therefore, the response transition found in this study reflects a cortical transformation and cortical encoding scheme change, as a function of only of FM dynamics.

The weak sideband performance in the upper sideband confusion matrix (Figure 5a) for $f_{FM}$ of 0.3 Hz and 0.5 Hz is probably due to narrowband noise at these two sideband frequencies (37.3 Hz and 37.5 Hz), which in turn gives lower signal-to-noise ratios at these points. Figure 4 shows the spectrum for one channel under all 9 stimulus conditions, and the narrowband noise at 37.3 Hz and 37.5 Hz can be clearly seen for all the stimulus conditions. The same reason accounts for the noisy distribution of encoding-type parameter $\alpha$ for $f_{FM}$ of 0.3 Hz because the phase calculated at this frequency point is also affected by noise.

For stimuli with faster-changing carriers ($f_{FM}$ up to 8 Hz), the upper sideband is consistently significant, which supports the use of modulation encoding by human auditory cortex to simultaneously represent the envelope and carrier dynamics.

To distinguish between types of modulation encoding used (e.g. AM encoding vs. PM encoding), we analyzed the distribution of the encoding-type parameter $\alpha$, which is approximately $\pi$ for pure PM encoding and approximately 0 or $2\pi$ for pure AM encoding (Figure 8). We found that for slower $f_{FM}$ stimuli (< 5 Hz, excluding the 0.3 and 0.5 Hz upper sideband), the encoding-type parameter $\alpha$ is approximately $\pi$ (Figure 7), indicating that those sidebands are due to the *phase* modulation of $f_{AM}$ by $f_{FM}$. In other words, the phase of the aSSR at $f_{AM}$ tracked the stimulus carrier frequency change, and because the carrier frequencies changed at certain frequencies ($f_{FM}$), the phase of $f_{AM}$ also changed at the corresponding $f_{FM}$ frequencies. These results for slower $f_{FM}$ were consistent with Patel & Balaban (2000) where the phase of the aSSR reliably tracked the carrier frequency contour of the tone sequences. There the carrier was a long, periodic, series of concatenated tone segments ($f_{FM}$ ~0.1 Hz), rather than the sinusoidally modulated carrier in our experiment. These results suggest that for stimuli with slow carrier dynamics ($f_{FM}$ < 5 Hz), auditory cortex tracks the carrier dynamics, i.e. the stimulus carrier frequency change, by modulating the phase of the aSSR at $f_{AM}$ accordingly.

As $f_{FM}$ increases, $\alpha$ begins to deviate from $\pi$ (Figure 7), indicating that encoding by phase tracking alone begins to deteriorate. Because upper sidebands are still present for those higher $f_{FM}$ stimuli (Figure 5a), modulation encoding (PM or AM or, e.g., both PM and AM) is still employed. One possibility is that another class of neurons have been recruited that use the amplitude, rather than the phase, of the aSSR at $f_{AM}$ to track the carrier dynamics. This kind of mechanism of AM encoding also elicits two sidebands around $f_{AM}$, but producing an encoding-type parameter $\alpha$ of approximately 0 (or $2\pi$), as shown in our simulation (Figure 8). We will explain this possibility in detail.

*Possible modulation coding schemes*

Patel & Balaban (2004) have proposed a model to explain their phase tracking results. They suggest that there are two groups of neurons, both of which fire in a phase-locked fashion to the envelope of the stimulus. One group of neurons tracks the carrier change by varying the firing phase within each $f_{AM}$ cycle, whereas the other group of neurons has only uniform random phase variation, although they still fire phase locked to $f_{AM}$ envelope. Using this model, the observed phase tracking results can be explained by reasonable neuronal mechanisms, specifically, the first group of neurons. This leads directly to responses dominated by PM encoding.

We propose another possible neural response type, the AM encoding neuron. These neurons also fire in a phase-locked fashion to the envelope of the stimulus ($f_{AM}$), but they change the *firing rate* rather than the *firing phase* within each cycle of $f_{AM}$ to track the carrier frequency change. Such kind of neuron group can elicit two sidebands around $f_{AM}$ with encoding-type parameter $\alpha$ around 0 (or $2\pi$).

The two proposed neuronal types are depicted in Figure 1b. In this illustrated example, the PM encoding neuron (third row) fires earlier for higher stimulus carrier frequency (first row) and fires later for lower stimulus carrier frequency (shown by the distance between the spike and the dotted line). In contrast, the AM encoding neuron (fourth row) in this example fires at a higher rate for higher stimulus carrier frequency and at a lower rate for lower stimulus carrier frequency.

Lu et al. (2001) found two largely distinct populations of neurons in auditory cortex of awake marmosets: one with stimulus-synchronized discharge (temporal code) coding for slow sound patterns, and the other using a rate code for rapidly repeating events. They suggest that the combination of temporal and rate codes provides a possible neural basis for wide range of temporal information representation in auditory cortex. Consistent with their suggestions, it is also possible that two groups of neurons, the PM encoding type and the AM encoding type neurons, are involved in encoding envelope and carrier dynamics simultaneously, and that the proportions depends on the stimulus dynamics. Single population models using both PM and AM are also possible and not ruled out by these results. For stimuli with low $f_{FM}$, more PM encoding type neurons are involved (temporal coding), and as $f_{FM}$ increases, more AM encoding type neurons begin to join, tracking carrier dynamics by AM (rate) coding.

MEG signals reflect combinations of responses from (potentially) many different neuronal classes. Therefore, when AM encoding neurons become involved in encoding stimulus dynamics, the observed MEG signals will be the sum of responses from both PM encoding type and AM encoding type neuronal responses. This affects the encoding-type parameter α distribution, as shown in the simulation results (Figure 8): the mixture of encoding populations causes the distribution to become more uniformly (broadly) distributed, rather than narrowly centered at π (for pure PM encoding). Saberi & Hafter (1995) proposed an FM-to-AM transduction hypothesis whereby a change in frequency is transmitted as a change in amplitude and suggested a common neural code (temporal code) for AM and FM sounds. In contrast, Moore & Sek (1996) suggest a two-stage FM sounds detection mechanism: the FM detection at low rate mainly depends on temporal information (phase locking to the carrier), whereas FM detection at higher rates (>10 Hz) depends mainly on changes in the excitation pattern (a "place" mechanism). Although both refer to pure FM detection, the ideas apply straightforwardly to our suggested interpretations.

In general, our results provide support for simultaneous encoding of envelope and carrier dynamics by modulation encoding in human auditory cortex. For stimuli with slow carrier dynamics (< 5 Hz), pure PM encoding is employed. For stimuli with faster carrier dynamics (here up to 8 Hz), modulation encoding is still present but probably not pure PM encoding. We propose the hypothesis that another group of neurons using AM encoding will be involved and continue to represent the stimulus dynamics. Importantly, our results provide natural hypotheses and predictions which can be tested in further neurophysiological studies.

## Acknowledgements

H.L., Y.W., and D.P. are supported by NIH R01 DC05660 to DP. We are grateful to Jeff Walker for his excellent technical assistance, and to 3 reviewers for thoughtful and detailed comments.

**Figure legends**

**Figure 1**

Modulation as an encoding method in engineering, and proposed neural mechanisms. a) A modulation signal modulates either the amplitude or the phase of the carrier signal to produce either an AM signal or a PM signal. Both signals produce a two-sideband pattern in spectrum (right). b) Possible neural modulation encoding mechanisms for AM encoding and PM encoding to simultaneously represent both stimulus carrier (first row, changes in stimulus carrier frequency) and stimulus envelope (second row, changes in stimulus amplitude) dynamics. A neuron employing PM encoding (third row) fires one spike per stimulus envelope cycle, as indicated by the dotted line, and the firing *phase* in each cycle depends on the instantaneous stimulus carrier frequency. A neuron employing AM encoding (the last row) changes firing rate according to the instantaneous stimulus carrier frequency, while keep the firing phase within each cycle fixed (aligned with the dotted line).

**Figure 2**

Stimulus examples. Top: the spectrograms of stimuli with $f_{FM}$ equal to 0.8 Hz and 2.1 Hz. The carrier frequency was modulated at a particular frequency (left, 0.8 Hz and right, 2.1 Hz), sinusoidally from 220 Hz to 880 Hz. Middle: the corresponding spectra of the stimulus examples in upper panel (left, 0.8 Hz and right, 2.1 Hz). Note that the spectra are broadband. Bottom, temporal waveform of stimulus with $f_{FM}$ equal to 2.1 Hz. The envelope of the stimulus is modulated sinusoidally at 37 Hz. Only one segment from 0.2 sec to 1.4 sec is shown to let the 37 Hz AM be seen more clearly. The carrier change can also be seen here. The stimuli have both dynamic envelope (lower panel) and carrier (upper panel).

**Figure 3**

Auditory steady state response (aSSR) at envelope modulation frequency (37 Hz). a) Spectrum of the response from one representative channel of one subject. The arrow indicates the evoked aSSR at 37 Hz. b) The phasor representation of aSSR at 37 Hz. It clearly shows a bilateral auditory MEG contour map. The arrow in each channel represents the Fourier coefficient at 37 Hz. The arrow length represents the magnitude and the arrow direction represents the phase. c) Grand average of dipole location for the aSSR at 37 Hz (red) and M100 (green), in axial, sagittal and coronal views. The two dipoles are localized in similar position of superior temporal cortex.

**Figure 4**

Spectrum and auditory-steady state response (aSSR) at sidebands at one channel in a representative subject. Each of the 9 figures represents the spectrum for each of the 9 different $f_{FM}$ stimulus conditions. The black arrow points to the aSSR at envelope modulation frequency (37 Hz) and can be observed for all the stimulus conditions. The grey arrows indicate the aSSR at corresponding upper sideband ($f_{AM} + f_{FM}$). For example, the stimulus with $f_{FM}$ of 0.3 Hz elicited 37.3 Hz aSSR (grey arrow). For this specific channel, all the stimulus conditions elicited corresponding upper sidebands except the stimulus with $f_{FM}$ of 5 Hz (grey arrow). (Subject R0458)

**Figure 5**

Empirical confusion matrix performance across 12 subjects. a) Upper sideband confusion matrix. b) Lower sideband confusion matrix. c) Total sideband confusion matrix. d) $f_{FM}$ confusion matrix. In the confusion matrix, each box represents the number of channels that elicited maximum magnitude at a particular sideband frequency (vertical axis) by a given stimulus condition (horizontal axis), for all stimulus conditions. Each row actually represents the histogram of best-driving stimulus for this specific sideband, and the sum of one row is equal to 120 (10 total channels × 12 subjects). If a particular sideband is significantly elicited by its corresponding stimulus, in one row (for one sideband frequency), the response on the diagonal will dominate the row. The peaked diagonal for all the 4 confusion matrixes can be seen here. The plot underlying each confusion matrix is the diagonal value plot normalized by channel number and subject number for the corresponding confusion matrix. For Figure 4a,b,d, the intensity reflects the response frequency's performance; since only one response frequency is tabulated, the range is from 0 (no aSSR at this frequency elicited at all) to 1 (aSSR elicited for all the 10 channels and all 12 subjects). For Figure 5c, since upper and lower sideband performances are summed, the range intensity is from 0 to 2 (two sideband frequencies elicited for all 10 channels and all 12 subjects). Generally, the diagonal values are above the threshold line, showing significant aSSR at these frequencies (sideband frequencies or $f_{FM}$ frequencies).

**Figure 6**

Simulated confusion matrix and sideband performance for a response using pure PM encoding with added Gaussian noise, assuming the same noise level and modulation index across all 9 conditions. The noise level is adjusted to match the empirical results. Sideband performance was extracted from the corresponding simulated confusion matrixes and shown below each confusion matrix. The dotted-starred line is the 95% threshold from a Monte Carlo simulation. a) Upper sideband confusion matrix. b) Lower sideband confusion matrix. c) Total sideband confusion matrix. The approximate match between empirical data (Figure 5a, 5b, 5c) and simulated data (Figure 6a, 6b, 6c) reflects modulation encoding.

**Figure 7**

Encoding-type parameter $\alpha$ performance. a) $\alpha$ histogram across the 10 selected channels and all 12 subjects for different stimulus conditions. The dotted line indicates the circular mean of $\alpha$. Both figures show that for stimulus with lower $f_{FM}$ (< 5 Hz), $\alpha$ is centered around $\pi$ and thus in the PM encoding region. When $f_{FM}$ becomes faster, $\alpha$ becomes more uniformly distributed and also leaves the PM encoding region. b) $\alpha$ plot for different stimulus condition. The mean of $\alpha$ is calculated using circular statistics. The standard error is calculated using bootstrap across all channels and subjects. The grey bar represents the PM encoding region (the middle grey bar, around $\pi$) and the AM encoding region (the upper and lower grey bar, around 0 or $2\pi$).

**Figure 8**

Simulation of α distributions using different proportions of neural AM and PM encoding. Neural AM and PM signals were created using same modulation (2 Hz) and carrier frequency (37 Hz). The neural signals were simulated as the sum of AM encoding (weight $\tau$) and PM encoding (weight $1-\tau$). α was calculated for each of the 1000 simulation trials and their distributions are shown for different $\tau$ (0.1 to 0.9 in step of 0.1). Small $\tau$ corresponds to dominant PM encoding, and large $\tau$ corresponds to dominant AM encoding. When $\tau$ α has a uniform-like distribution.

**Figure 1**

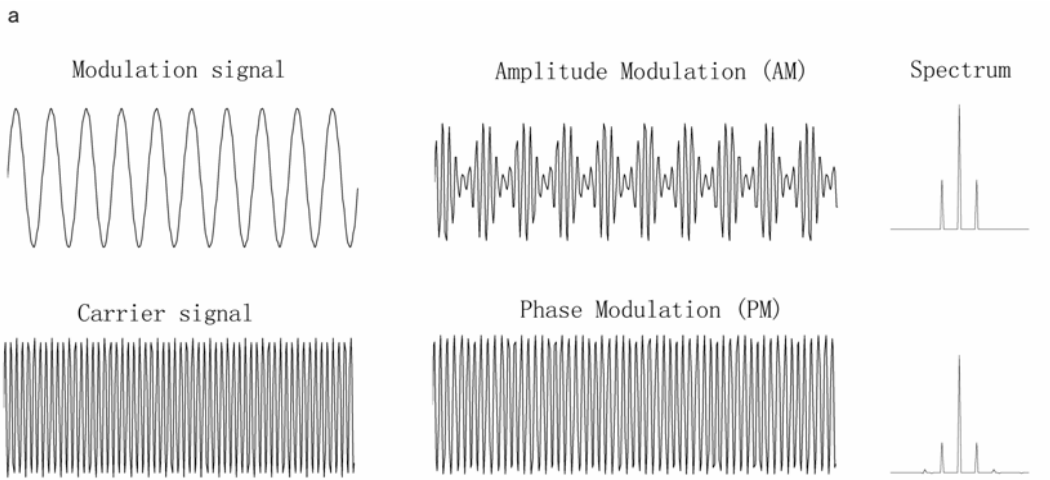

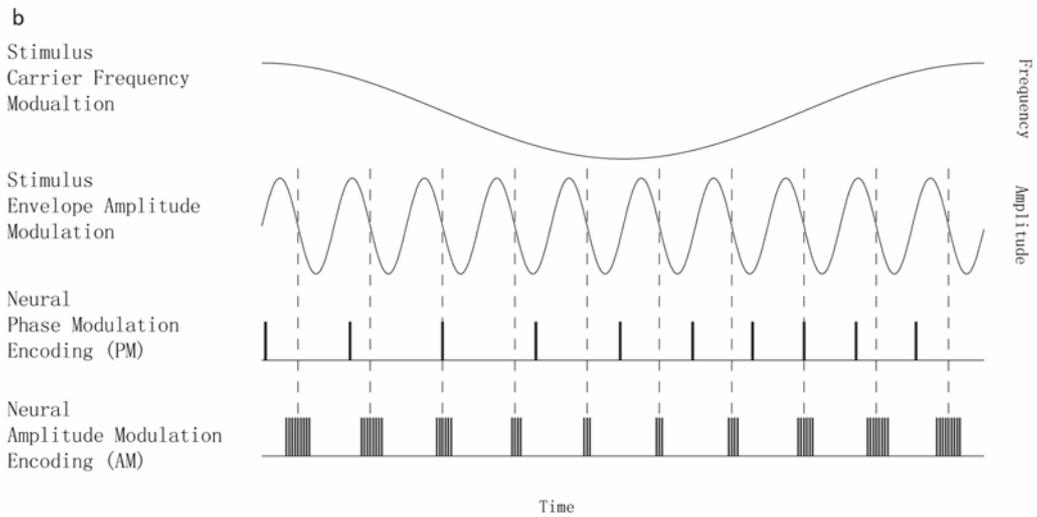

                                   2/27/06

**Figure 2**

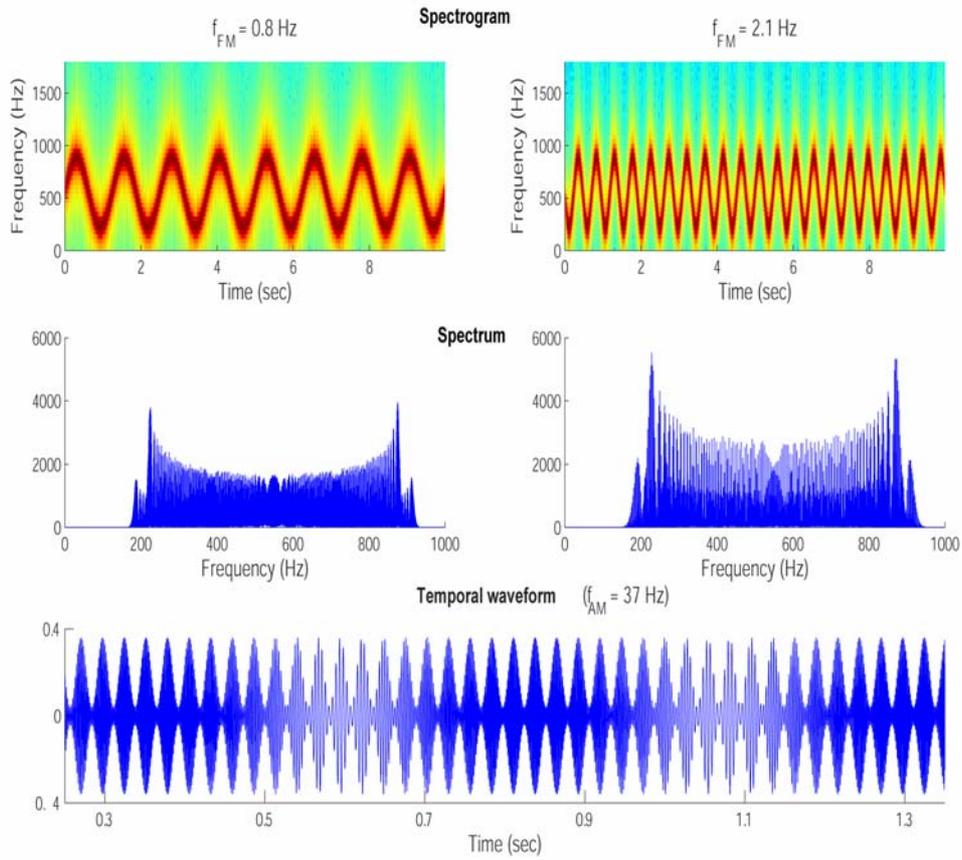

**Figure 3**

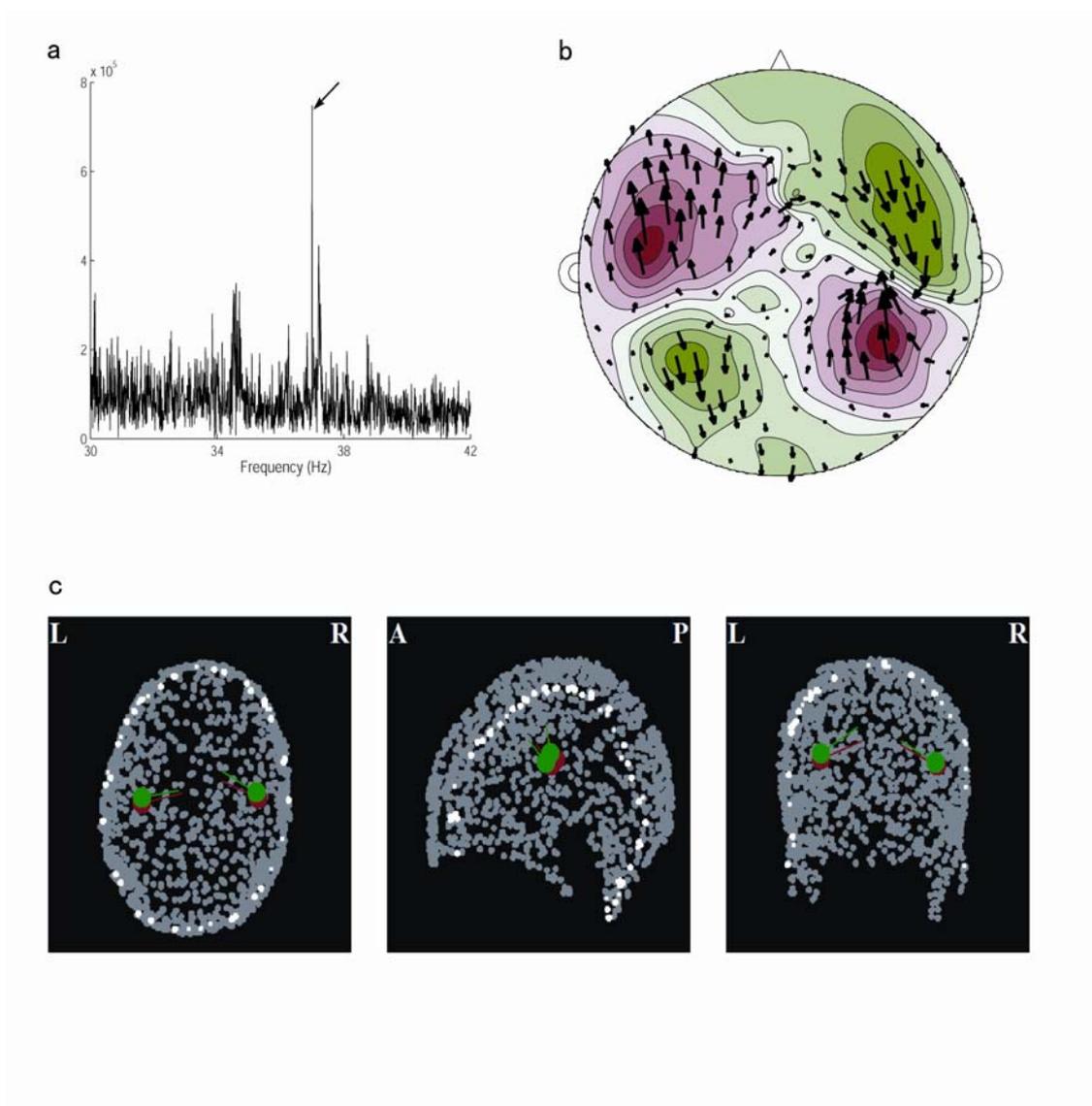

**Figure 4**

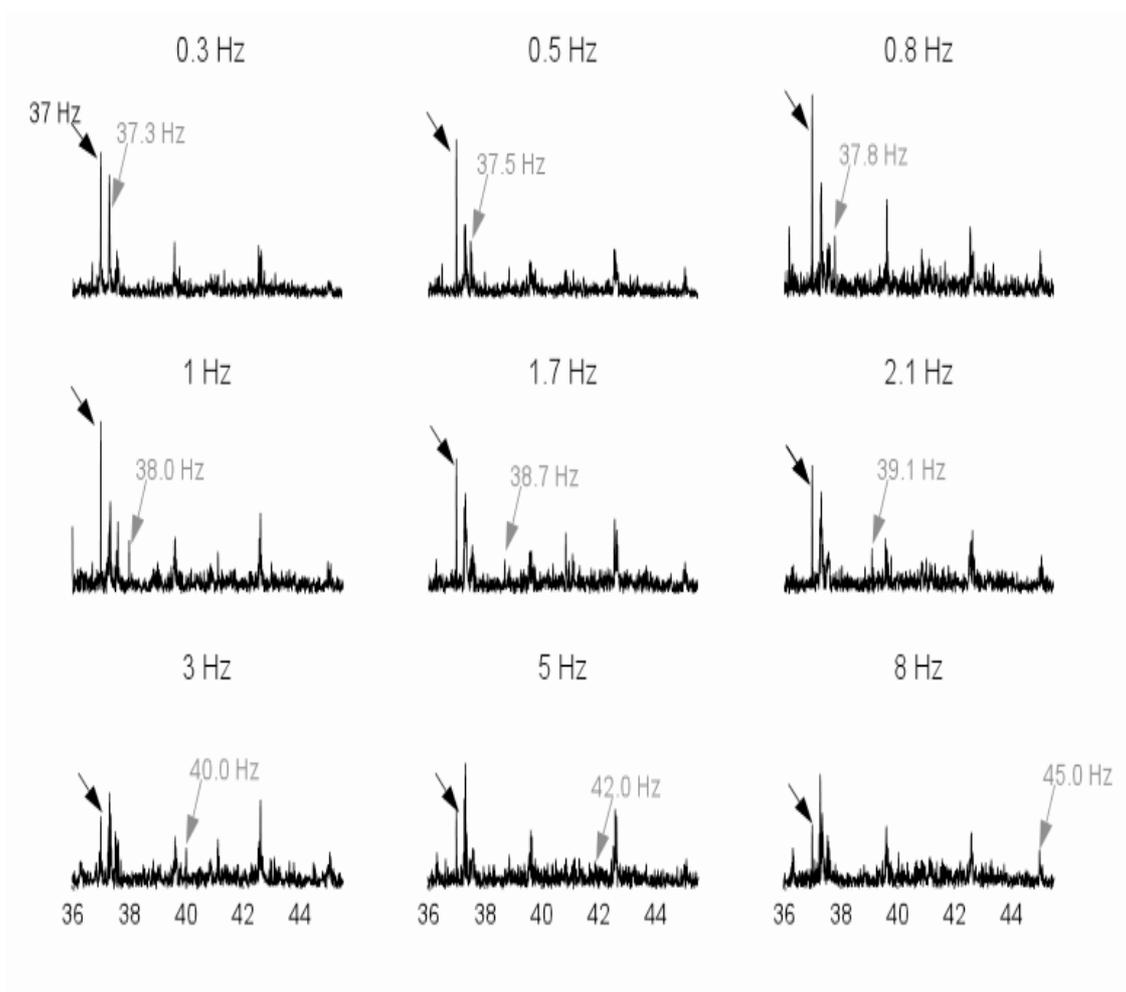

**Figure 5**

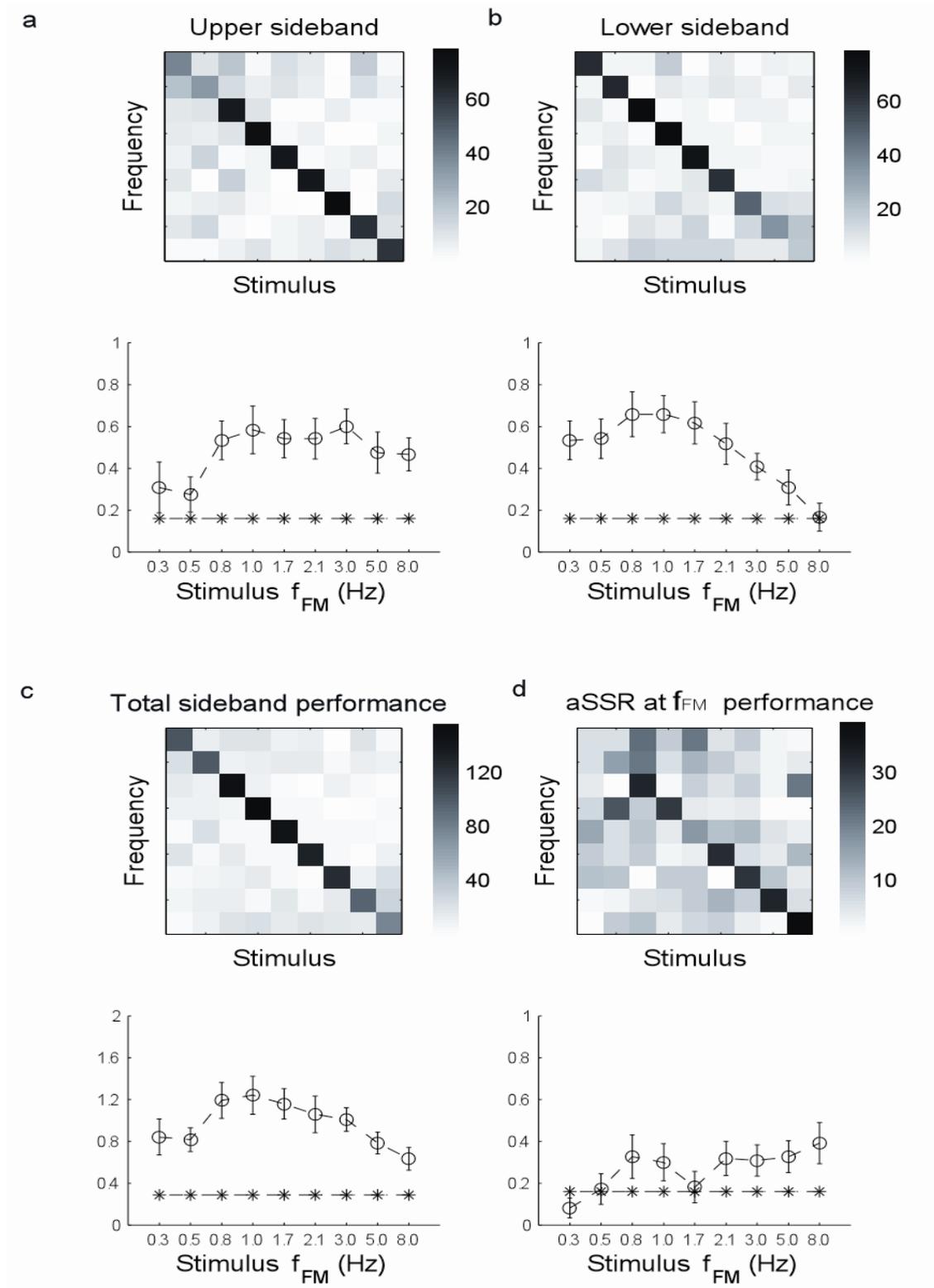

**Figure 6**

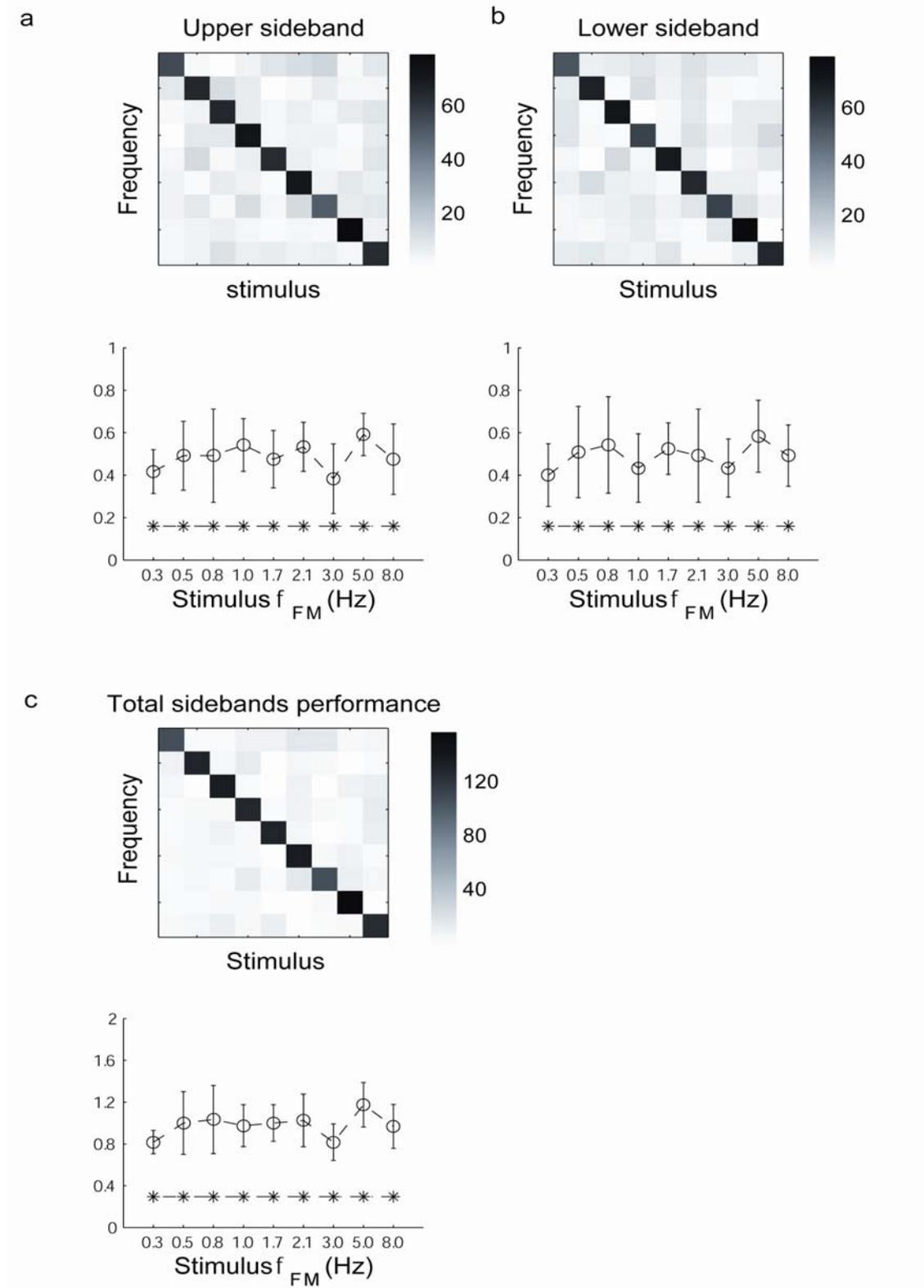

**Figure 7**

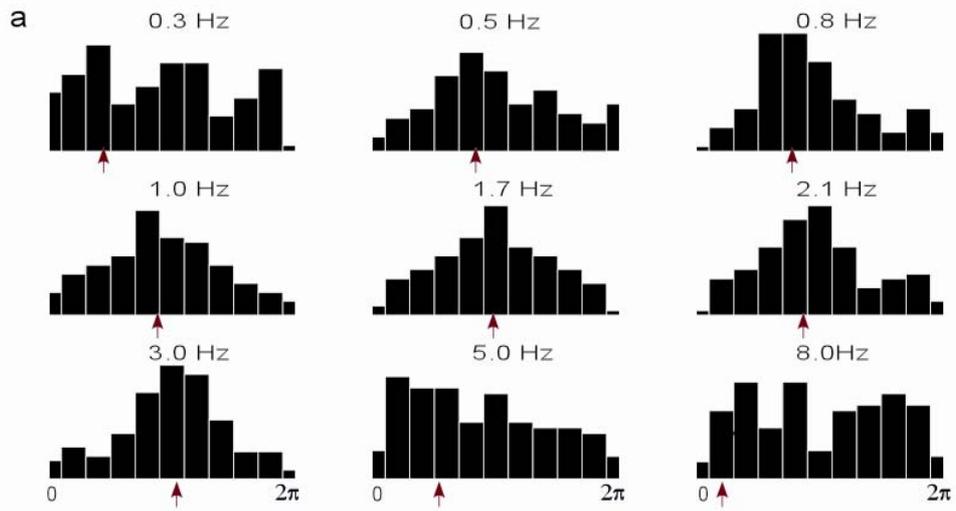

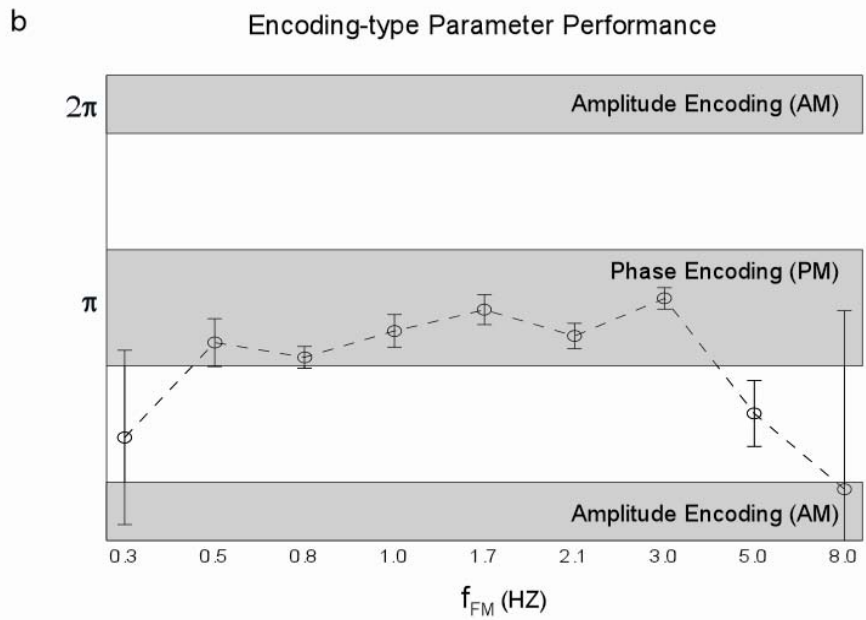

**Figure 8**

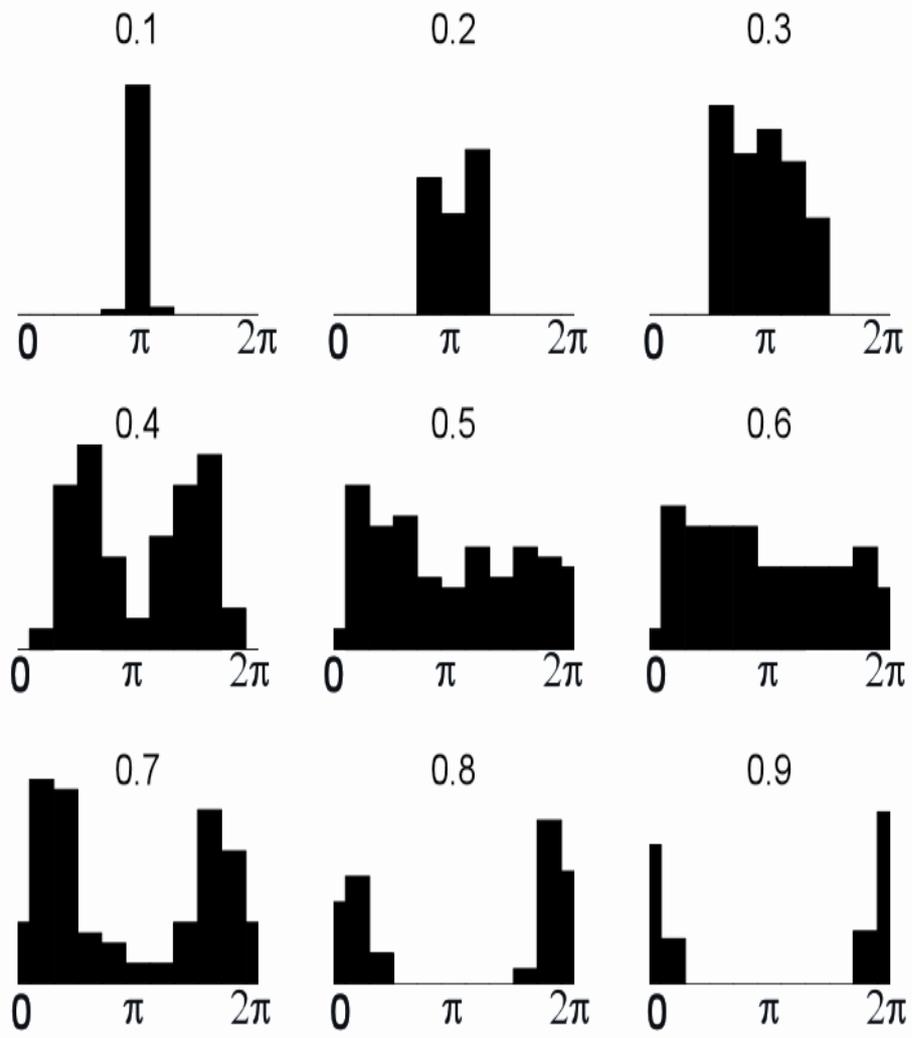

 2/27/06